\documentclass[11pt,a4]{article}
\usepackage{hyperref}
\usepackage{lineno}
\usepackage[affil-sl]{authblk}
\usepackage{setspace}
\usepackage{geometry}
\usepackage{graphicx}
\usepackage{subcaption}
\usepackage{bm}
\usepackage{multicol}
\usepackage{amssymb}
\usepackage{mathtools}
\usepackage{slashed}
\usepackage{cite}
\usepackage{authblk}

\newgeometry{vmargin={20mm}, hmargin={15mm,20mm}}   
\hypersetup{colorlinks=true,allcolors=blue} 
\title{Letter of Intent: Search for sub-millicharged particles at J-PARC}
\author[1]{Suyong Choi}
\author[1]{Jeong Hwa Kim}
\author[1]{Eunil Won}
\author[1]{Jae Hyeok Yoo}
\affil[1]{Korea University, Seoul, Korea}
\author[2]{Matthew Citron}
\author[2]{David Stuart}
\affil[2]{University of California, Santa Barbara, California, USA}
\author[3]{Christopher S. Hill}
\affil[3]{The Ohio State University, Columbus, Ohio, USA}
\author[4]{Andy Haas}
\affil[4]{New York University, New York, New York, USA}
\author[5]{Jihad Sahili}
\author[5]{Haitham Zaraket}
\affil[5]{Lebanese University, Hadeth-Beirut, Lebanon}
\author[6]{A. De Roeck}
\author[6]{Martin Gastal}
\affil[6]{CERN, Geneva, Switzerland}

\date{}
 
\begin{document} 
\maketitle

\begin{abstract}
We propose a new experiment sensitive to the detection of millicharged particles produced at the $30$ GeV proton fixed-target collisions at J-PARC. The potential site for the experiment is B2 of the Neutrino Monitor building, $280$ m away from the target. With $\textrm{N}_\textrm{POT}=10^{22}$, the experiment can provide 	sensitivity to particles with electric charge $3\times10^{-4}\,e$ for mass less than $0.2$ $\textrm{GeV}/\textrm{c}^2$ and $1.5\times10^{-3}\,e$ for mass less than $1.6$ $\textrm{GeV}/\textrm{c}^2$. This brings a substantial extension to the current constraints on the charge and the mass of such particles.

\end{abstract}


\newpage 

\section{Motivation for the experiment}


%



Electric charge quantization is a long-standing question in particle physics. While fractionally charged particles (millicharged particles hereafter) have typically been thought to preclude the possibility for Grand Unified Theories (GUTs), well-motivated dark-sector models have been proposed to predict the existence of millicharged particles while preserving the possibility for unification. Such models can contain a rich internal structure, providing candidate particles for dark matter. One well motivated mechanism that leads to millicharged particles is to introduce a new $U(1)$ in dark sector with a massless dark-photon ($A'$) and a massive dark-fermion ($\chi$)~\cite{Holdom:1985ag}, 
\begin{eqnarray}
\mathcal{L}_{\textrm{dark sector}} 
= -\frac{1}{4} A'_{\mu\nu} A'^{\mu\nu} + i \bar{\chi} \left( \slashed{\partial} + i e' \slashed{A}' + i \textrm{m}_{\chi}\right) \chi 
  - \frac{\kappa}{2} A'_{\mu\nu} B^{\mu\nu}
\end{eqnarray}
where the last term shows that $A'$ and $B$ kinetically mix with the mixing parameter $\kappa$. After replacing $\slashed{A}'$ with $\slashed{A}' + \kappa B_\mu$, the coupling between the dark fermion and $B$ becomes $\kappa e'$ ($~\kappa e' \bar{\chi} \slashed{B} \chi$).  The charge of $\chi$ varies by the size of mixing, so $\chi$ can be a millicharged particle. Hereafter, $\chi$ is used to denote millicharged particles.

A number of experiments have searched for millicharged particles, including in an electron fixed-target experiment~\cite{MilliQ}, proton-proton colliders~\cite{Chatrchyan_2013,Chatrchyan_2013_2,Ball:2020dnx}, and neutrino experiments~\cite{Davidson:1991si}. A comprehensive review is in Reference~\cite{Emken_2019}. In the parameter space of the charge ($Q$) and the mass ($m_\chi$), the region of $m_\chi>0.1$ $\textrm{GeV}/\textrm{c}^2$ and $Q<10^{-2}e$ is largely unexplored.

Proton fixed-target experiments provide a solid testing ground for $\chi$s. The particle flux is much larger than the collider experiments and they can reach higher energy regime than electron fixed-target experiments. The sensitivity of such experiments to $\chi$s can reach beyond $Q_\chi \sim 10^{-3}e$ for a wide mass range from a few $\textrm{MeV}/\textrm{c}^2$ to a couple of $\textrm{GeV}/\textrm{c}^2$. We propose an experiment, SUBMET (SUB-Millicharge ExperimenT), which utilizes the $30$ GeV J-PARC proton beam to search for $\chi$s in this unexplored region. 

\section{Production of millicharged particles at J-PARC}

\begin{figure}[h]
\begin{center}
\includegraphics[width=0.7\linewidth]{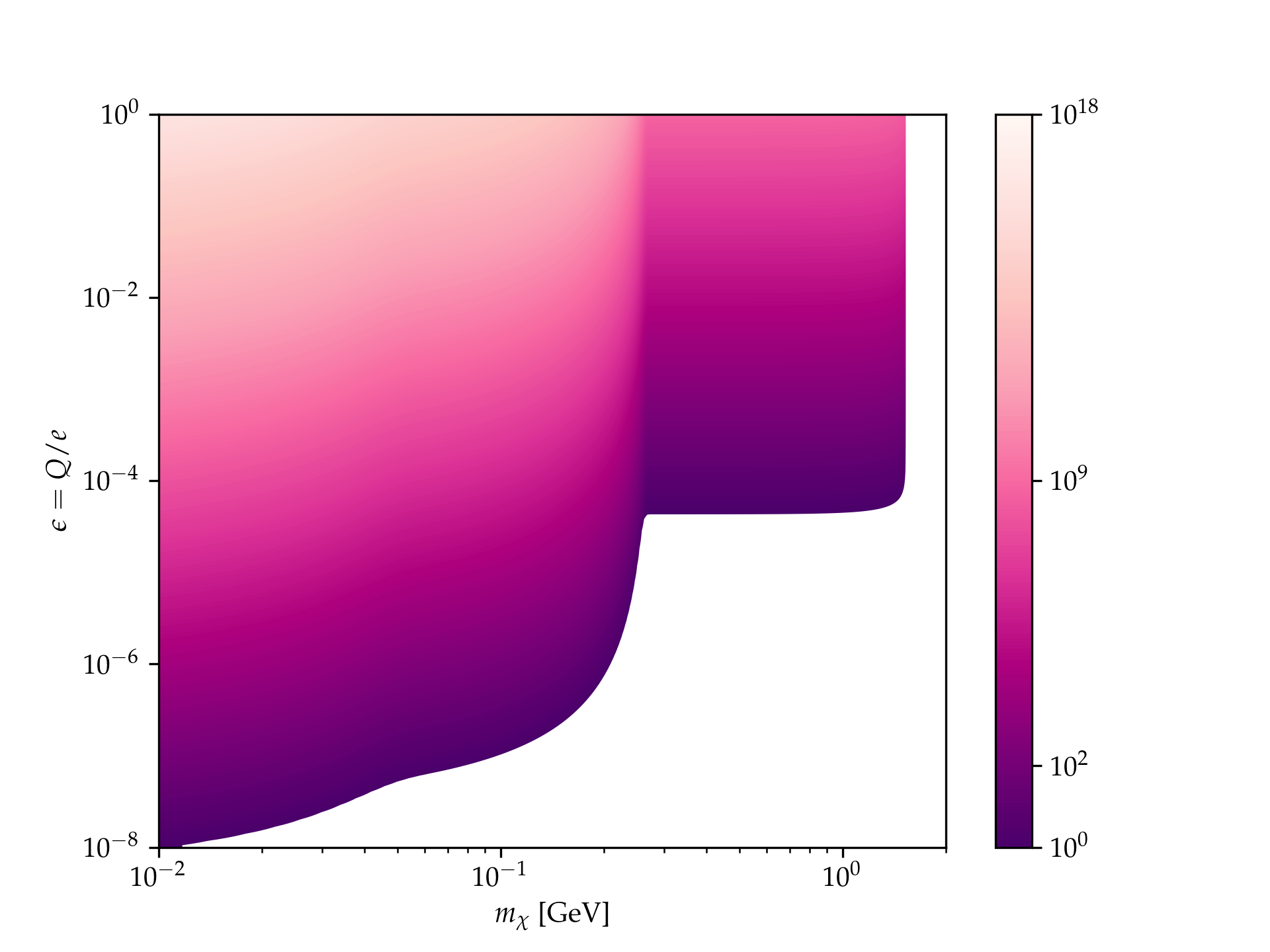}
\end{center}
\caption{Expected number of $\chi$s that reach a $0.5 \textrm{ m}^2$ detector located at 280 m from the target. $\textrm{N}_\textrm{POT}=10^{22}$ is assumed.}
\label{fig:nsig_acc}
\end{figure}

We consider $\chi$s with charge $Q$ produced from $\pi^0, \eta$ and $J/\psi$ neutral meson decays. The $\Upsilon$ production is irrelevant because the center-of-mass energy $\sqrt{s}$ is $7.5$ GeV for the collisions between the $30$ GeV proton beam and the fixed target. The lighter mesons ($\mathfrak{m} = \pi^0, \eta$) decay through photons ($\pi^0, \eta \to \gamma\chi\bar{\chi}$), while $J/\psi$ mostly decay to $\chi$s directly ($J/\psi \to \chi\bar{\chi}$). In both cases, $m_\chi$ up to $m_\mathfrak{m}/2$ is kinematically allowed. The number of produced $\chi$s $N_\chi$ can be calculated by the equation in~\cite{fermini}, 
\begin{eqnarray}
N_\chi \propto c_\mathfrak{m} \epsilon^2 \textrm{N}_\textrm{POT} \times f(\frac{m^2_\chi}{m^2_\mathfrak{m}})
\end{eqnarray}
where $c_\mathfrak{m}$ is the number of mesons produced per proton-on-target (POT), $\textrm{N}_\textrm{POT}$ is the total number of POT, $\epsilon = Q/e$, and $f$ is a phase space related integral that goes to $0$ as $m_\chi$ approaches $m_\mathfrak{m}/2$. Figure~\ref{fig:nsig_acc} shows the expected number of $\chi$s that reach the detector with face area of $0.5~\textrm{m}^2$, located $280$ m from the target. Assuming $\textrm{N}_\textrm{POT}=10^{22}$ that corresponds to running the experiment for $\sim3$ years, the expected number of $\chi$s is in the order of $10^{16}$ at $\epsilon=1$ and it is in the order of $10^{9}$ at $\epsilon=10^{-4}$.

\section{Site and detector concept}



In J-PARC, a $30$ GeV proton beam is incident on a graphite target to produce hadrons that subsequently decay to a muon and muon neutrino in the decay volume. The remaining hadrons are then dumped in the beam dump facility. Since they are MIPs, muons penetrate the beam dump and are identified by the muon monitor located behind the beam dump facility. The on-axis near detector (INGRID) is inside the Neutrino Monitor (NM) building located at $280$ m from the target. The space between the muon monitor and INGRID is filled with sand. 

\begin{figure}[h]
\begin{center}
\includegraphics[width=0.99\linewidth]{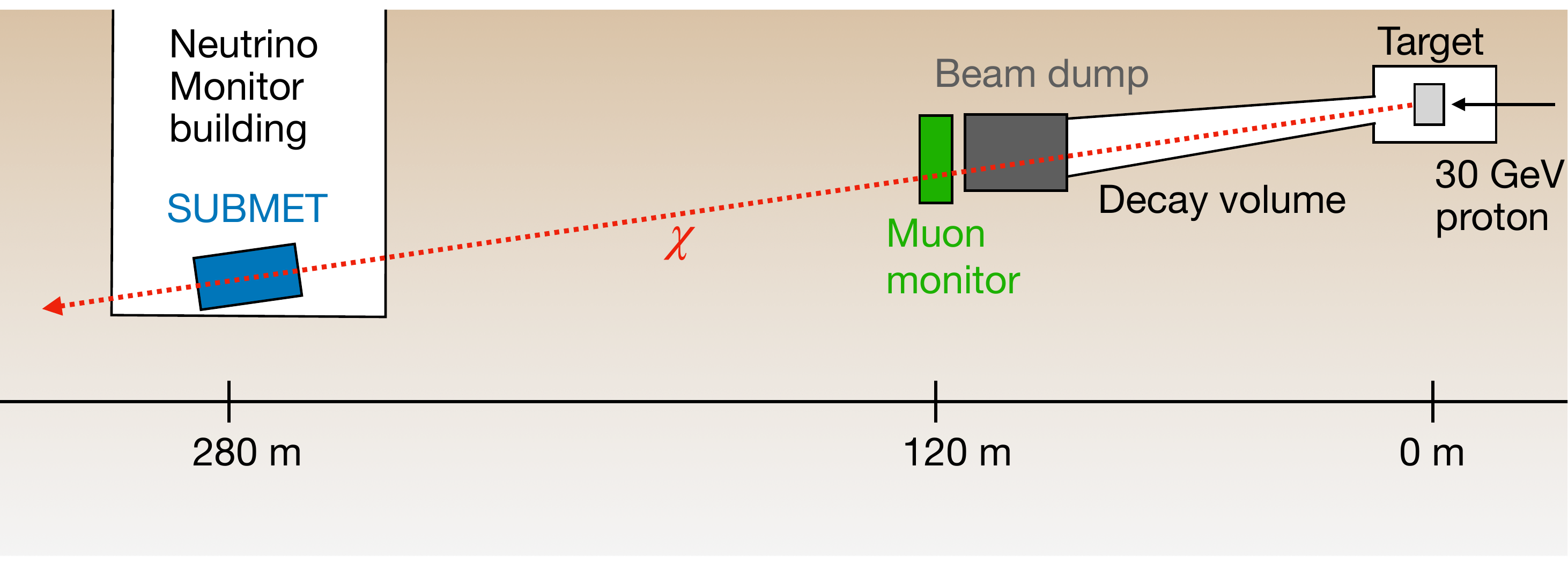}
\end{center} 
\caption{Illustration of the experimental site. $\chi$s are produced near the target and reach SUBMET after penetrating the beam dump, the muon monitor and the sand.}
\label{fig:site}
\end{figure}

If $\chi$s are produced, because of their low interaction probability, they will penetrate the beam dump facility, the muon monitor, and the sand without a significant energy loss. Therefore, they can be detected at the NM building if a detector sensitive to identifying such particles is installed. We consider the area behind the V-INGRID on B2 ($\sim30$ m underground) as a potential detector site. The distance from the axis of the neutrino beam is $\sim 5$ m. As described below, the signal acceptance is only slightly smaller than the on-axis location. 

\begin{figure}[h]
\begin{center}
\includegraphics[width=0.99\linewidth]{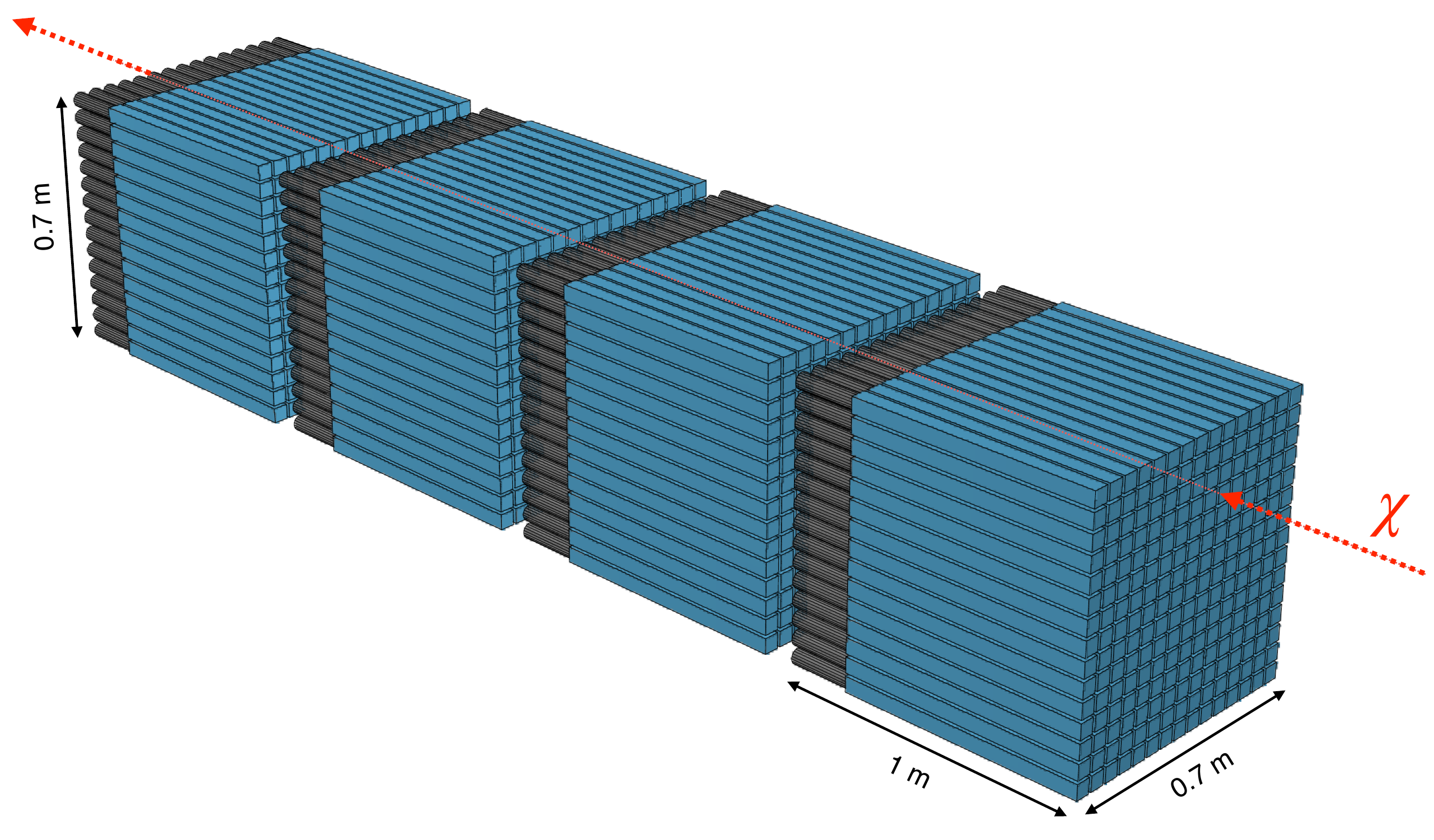}
\end{center}
\caption{Demonstration of the SUBMET detector. There are four layers of scintillator bars (blue). At one end of each bar a PMT (black) is attached A $\chi$ will penetrate 4 layers in a narrow time window.}
\label{fig:detector}
\end{figure}

The detector concept proposed here is based on a similar proposal made in~\cite{Ball:2016zrp}, namely, using a segmented detector with large scintillator bars as the core detector technique. In order to be sensitive to charges down to $10^{-3}e$, a thick sensitive volume is needed. It is advantageous to segment the large volume because it reduces backgrounds due to dark currents and shower particles from cosmogenic muons to a negligible level. It also allows for utilizing the directionality of the incident $\chi$s to further suppress non-pointing particles. The detector, as shown in Figure~\ref{fig:detector}, is composed of 4 layers of stacked $5\times5\times80$ $\textrm{cm}^3$ plastic scintillator bars. They are aligned such that the produced $\chi$s pass through all layers in a narrow time window. In each layer there are $14 \times 14$ scintillator bars, so the area of the detector face is about $0.5 \textrm{ m}^2$. A prototype of a detector with a similar design has been installed at the LHC, and shown robustness and sensitivity to $\chi$s~\cite{Ball:2020dnx}.  

At the end of each scintillator bar a photodetector is attached to convert the photons to an electronic signal. There are several options, but we consider Photomultipliers (PMTs), because of their large area coverage, low cost, and low dark current. Since the size of a bar is $5\times5$ $\textrm{cm}^2$, a fast commercial $50$ mm diameter PMT is suitable. Including PMTs, the total volume of the detector is approximately $0.7\times0.7\times4$ $\textrm{m}^3$. 

\begin{figure}[h]
\begin{center}
\includegraphics[width=0.8\linewidth]{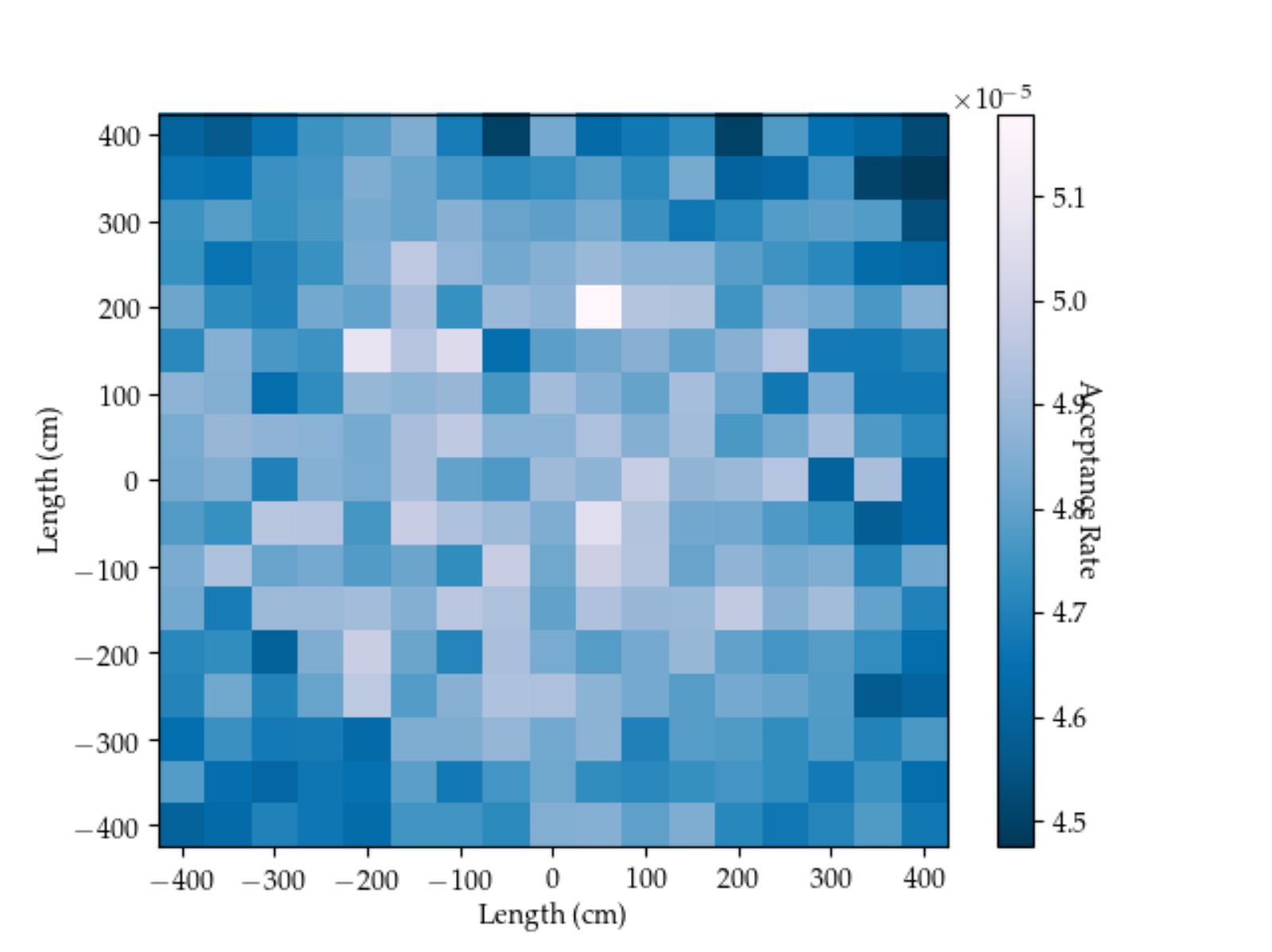}
\end{center}
\caption{Signal acceptance rate at 280 m from the target. One box corresponds to $0.5\times0.5$ $\textrm{m}^2$.}
\label{fig:acc}
\end{figure}

The signal acceptance rate, the fraction of $\chi$s that go into the detector area of $0.5 \textrm{ m}^2$ at $280$ m from the target, is calculated. Since the photons are massless, the direction of the photons and that of $\chi$s should be aligned. Therefore, the direction of photons from $\pi^0$'s is used as a proxy for the direction of $\chi$s. Figure~\ref{fig:acc} shows the acceptance rate in $8\times8$ $\textrm{m}^2$ plane where the axis of the neutrino beam is at the center. The acceptance rate does not depend on the location within this range strongly, \textit{i.e.}, near the corners the rate is only $\sim 10$\% lower than the central region because of the large distance from the target. This provides some flexibility in selecting the location of the detector.     

The operation of SUBMET does not interfere with the operation of the neutrino program and dedicated beam time will not be required. The detector is located behind the V-INGRID detector, so we don't expect any interference with the measurements of the neutrino beam properties and the operation of the INGRID detector. The detector will be triggered by its own trigger system and the experiment will have an independent data stream. Therefore, the operation of this new experiment does not have any conflict with the existing programs at J-PARC.

\section{Background sources}



The $\chi$s that reach the detector will go through all layers within a $\sim20$ ns time window (coincidence signal). In this section, the background sources that can mimic this coincidence signal are discussed. They can be divided into three categories, random coincidence, beam-induced, and cosmic-induced backgrounds.

In PMTs, spurious current pulses can be produced by thermal electrons liberated from the photocathode. If the rate of such pulses (dark count rate, DCR) is high, the rate of random coincidence can be substantially large even if the time window for coincidence signal is $20$ ns. The random coincidence rate can be calculated by $R = n N^n \tau^{n-1}$ where $n$ is the number of layers, $N$ is the DCR, and $\tau$ is the coincidence time window. Using a typical DCR of 
$500$ Hz, $n=4$, and $\tau=20$ ns, the random coincidence rate is $6 \times 10^{-5}$ per year. Since there can be $14\times14=196$ such coincidence signals, the total coincidence rate is $\sim 0.01$ per year. 

Muons are produced from the pion decays in the decay volume together with neutrinos. The density of quartz, which typically takes up the largest fraction of sand, is $2.65$ $\textrm{g}/\textrm{cm}^3$ and $dE/dx = 1.699 \textrm{ MeV}/\textrm{cm}^2 \textrm{g}$~\cite{Groom:2001kq}, so the energy loss of a MIP in $>100$ m of sand is much larger than $30$ GeV. Therefore, such beam-induced muons can't reach the detector. Although the muons from the pion decays can't reach the detector, neutrinos can and may interact with the scintillator material to produce small signals. The number of neutrino interaction events in INGRID is $\sim 1.5 \times 10^8$ for $\textrm{N}_\textrm{POT}=10^{22}$~\cite{Abe:2011xv}. Since a large fraction of INGRID material is iron, we can use the rate of INGRID as a upper bound for SUBMET. One layer of SUBMET is approximately $30$ times smaller, so the rate is $\sim 5 \times 10^6$ for $\textrm{N}_\textrm{POT}=10^{22}$ in one layer of SUBMET. Requiring coincidence in 4 layers, the expected number of this background is negligible. The interaction of the neutrinos and the material of the wall of the NM building in front of the detector can produce muons that go through the detector. These muons can be identified and rejected by installing scintillator plates between the wall and the detector or by using the very large scintillation yield of a muon that can be separated from millicharge signal. 

Cosmic muons that penetrate the cavern or the materials above the detector can produce a shower of particles that is large enough to hit all layers simultaneously. In such events, the hits in multiple layers can be within the coincidence time window and will look like a signal event. The particles in the shower generate more photons than $\chi$s, so the signals from cosmic muon showers can be rejected by vetoing large pulses. As done to tag the muons produced in the wall of the NM building in front of the detector, scintillator plates can be installed covering the whole detector to tag any ordinary-charge particles or photons incident from top and sides of the detector. These auxiliary components were proven to be  effective in rejecting events with such particles~\cite{Ball:2020dnx}. In addition, the cosmic shower penetrates the detector sideways, leaving hits in multiple bars in the same layer, while $\chi$s will cause a smaller number of hits. A cosmic shower and signals form radioactive decays overlapping with dark current can be another source of background. Since the rate of this background depends on the environment strongly, a precise measurement can be performed \textit{in situ} only. 

To estimate the sensitivity of the experiment, we assume that the total background over three years of running is $O(1)$ event.




\section{Sensitivity}



\begin{figure}[ht]
\begin{center}
  \includegraphics[width=0.75\linewidth]{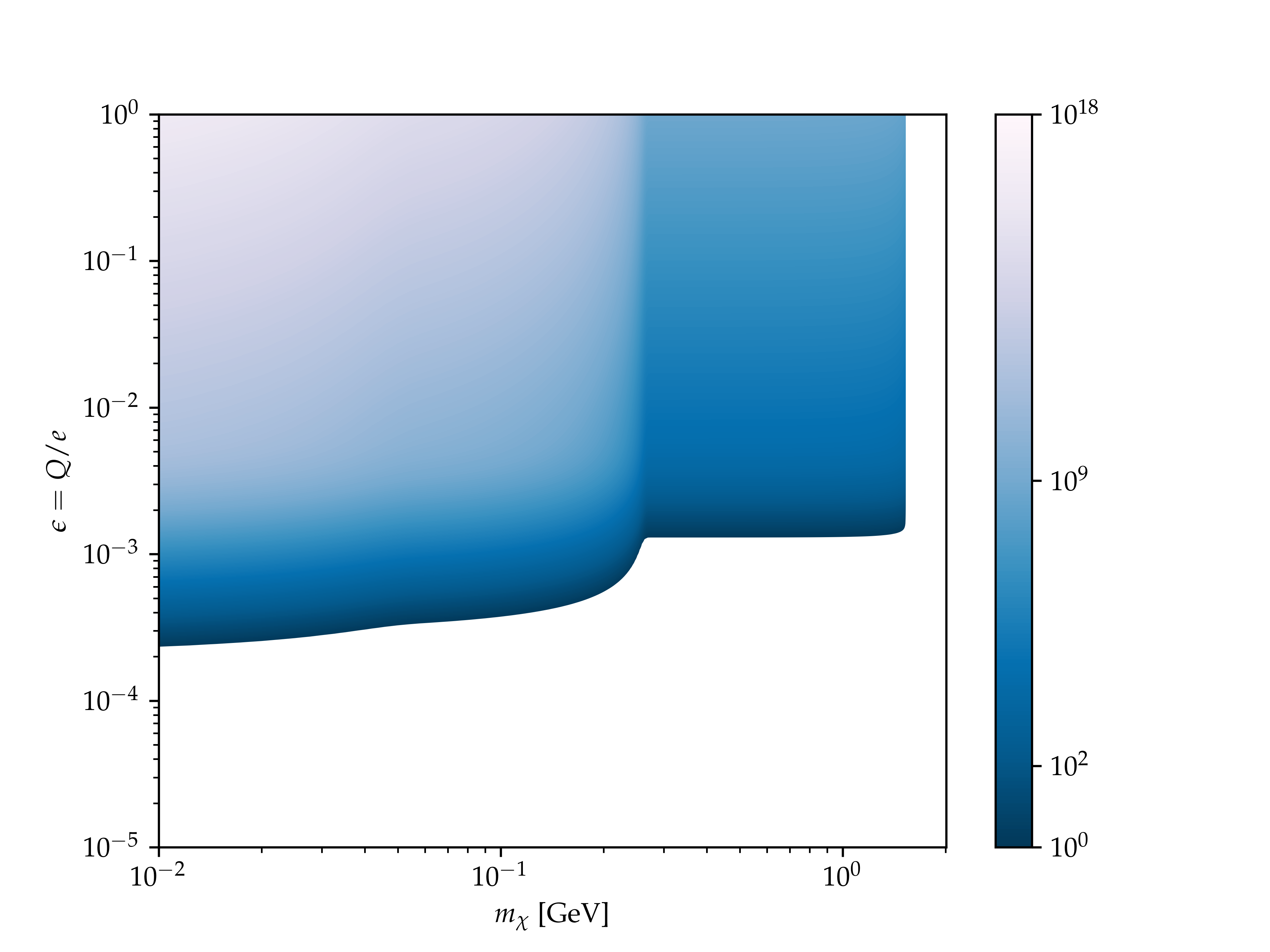}
\end{center}
\caption{Expected number of detected signal events $s$ in $\textrm{N}_\textrm{POT}=10^{22}$. 
There is a sharp drop in $\epsilon<10^{-3}$ due to small $\textrm{N}_\textrm{PE}$.}
\label{fig:nsig_obs}
\end{figure}

The probability of detecting a $\chi$ in a $n$-layer detector is given by Poisson distribution $P = (1-e^{-\textrm{N}_\textrm{PE}})^n$ where $\textrm{N}_\textrm{PE}$ is the number of photoelectrons. $\textrm{N}_\textrm{PE}$ is proportional to the quantum efficiency (QE) of PMT, $\epsilon^2$, and the number of photons that reach the end of the scintillator ($N_\gamma$). The $\epsilon^2$ term comes from the fact that the energy loss of a charged particle in matter is proportional to $\epsilon^2$. In order to get $N_\gamma$ a \texttt{GEANT4}~\cite{Agostinelli:2002hh} simulation is performed. Using a $5\times5\times80$ $\textrm{ cm}^3$ Saint-Gobain BC-408 plastic scintillator with surface reflectivity of 98\%, we get $N_\gamma =  6.3 \times 10^5$. Using $\textrm{QE}=30$\% and taking the area of the PMT into account, we get $\textrm{N}_\textrm{PE}=1.3\times 10^5 \epsilon^2$. We can now calculate the total number of signal events $s = N_\chi P$ measured by the detector. Figure~\ref{fig:nsig_obs} shows the number of detected signal events $s$ in the plane of $\epsilon$ and $m_\chi$. 

\begin{figure}[h]
\begin{center}
\includegraphics[width=0.99\linewidth]{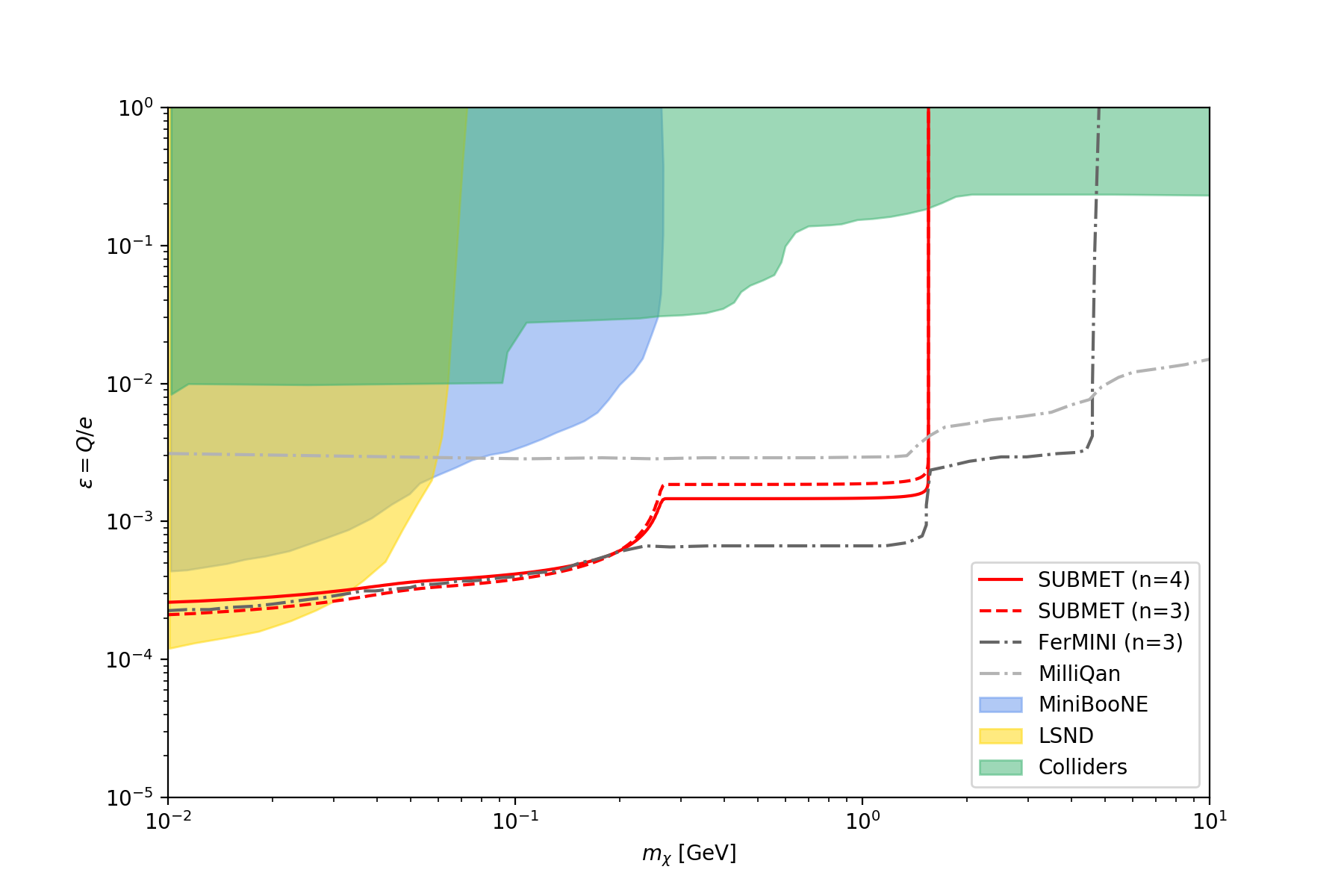}
\end{center}
\caption{Exclusion at $95$\% CL for $\textrm{N}_\textrm{POT}=10^{22}$. For comparison, the constraints from previous experiments are shown as shaded area and the expected sensitivity of the proposed experiments are drawn in gray dotted line. Note that FerMINI curve is from 3-layer detector design.}
\label{fig:limit}
\end{figure}

Figure~\ref{fig:limit} shows the $95$\% CL exclusion curve for $\textrm{N}_\textrm{POT}=10^{22}$. SUBMET provides exclusion down to $\epsilon=3\times10^{-4}$ in $m_\chi<0.2$ $\textrm{GeV}/\textrm{c}^2$ and $\epsilon=1.5\times10^{-3}$ in $m_\chi<1.6$ $\textrm{GeV}/\textrm{c}^2$. We do not consider a systematic uncertainty on $b$ because it does not have a significant impact on the exclusion limit (Table~\ref{tab:det_config}).  
Sensitivity of SUBMET is comparable with that of FerMINI in $m_\chi<0.2$ $\textrm{GeV}/\textrm{c}^2$. Above that mass range the performance of SUBMET is worse because the production rate of $J/\psi$ is much lower with the $30$ GeV proton beam.

As shown in Fig~\ref{fig:nsig_obs}, the number of signal events recorded by the detector drops rapidly in $\epsilon<10^{-3}$ due to small $\textrm{N}_\textrm{PE}$. Therefore, one can expect that increasing $\textrm{N}_\textrm{PE}$ or $N_\chi$ would not have a large impact on the sensitivity. This will be discussed in a quantitative way in the next section.


\section{Alternative detector design}

In this section, we discuss how sensitivity to sub-millicharged particles depends on the configuration of the detector. If further optimization of the detector is needed, we can use this quantitative study as a guidance. 

Silicon Photomultipliers (SiPMs) can be considered instead of PMTs because of its excellent sensitivity to single photons and small size. A challenge is that they have a typical DCR of $5 \textrm{ MHz} / \textrm{cm}^2$ at room temperature, about three orders of magnitude larger than that of PMTs. This leads to a significantly larger random coincidence rate. Since the dark current is a thermal effect, it has a strong dependence on temperature. Typical reduction rate is a factor of $\sim2$ per $10^{\,\circ}$C. So, to suppress the DCR to the similar level of PMT's, the temperature should be lowered by of order of $100^{\,\circ}$C. In the scenario where SiPMs are used, the size of scintillator bars is chosen to be $1\times1$ cm due to difficulty in covering the large area of the scintillator. In this case the light attenuation effect increases due to more reflections and the photon yields decreases by a factor of $2$.   

Searches in the sub-millicharge regime rely on the Poissonian fluctuation of small $\textrm{N}_\textrm{PE}$. Approximating $P \simeq \textrm{N}_\textrm{PE}$ to the first order for small $\epsilon$, we arrive at the following relation 
\begin{eqnarray}
\textrm{N}_{\epsilon=1,\chi} \epsilon^2 (\textrm{N}_{\epsilon=1,\textrm{PE}} \epsilon^2)^n \geq s^{\textrm{95\%}}
\end{eqnarray}
where the subscript $\epsilon = 1$ refers to the values at $\epsilon = 1$ and $s^{\textrm{95\%}}$ is the number of signal events that provides 95\% exclusion limit. Reordering in terms of $\epsilon$, we recognize the exclusion limit at 95\% CL is
\begin{eqnarray}
\label{eq:eps95}
\epsilon = \left( \frac{\textrm{N}_{\epsilon=1, \textrm{PE}}^n \textrm{N}_{\epsilon=1,\chi}}{s^{\textrm{95\%}}} \right)^{-\frac{1}{2n+2}}.
\end{eqnarray}
Not surprisingly, the $\epsilon$-power term induces a sharp cutoff in signal at $\epsilon \sim O(10^{-4})$ as shown in Figure~\ref{fig:nsig_obs}. This limits the sensitivity to the same regime regardless of the detector configuration. 

\begin{table}
\begin{center}
 \begin{tabular}{c c c | c} 
 \hline
 $N_\chi$(relative)  &  $\textrm{N}_\textrm{PE}$(relative) & $b$ &  Exclusion limit on $\epsilon$ for $m_{\chi} = 10$ $\textrm{MeV}/\textrm{c}^2$\\ 
 \hline\hline
 1 & 1 & 1        & $2.6\times10^{-4}$\\ 
  \hline
 2 & 1 & 1        & $2.4\times10^{-4}$\\
 \hline
 1 & 1 & 100      & $3.0\times10^{-4}$\\
 \hline   
 1 & 2 & 1        & $1.9\times10^{-4}$\\
 \hline
 1 & 3 & 1        & $1.7\times10^{-4}$\\
 \hline
\end{tabular}
\end{center}
\caption{Various detector configurations and their sensitivity. $N_\chi$ (relative) is the yields of signal events within the acceptance relative to the baseline, $\textrm{N}_\textrm{PE}$ (relative) is the number of photon electrons relative to the baseline $\textrm{N}_\textrm{PE}=1.3\times10^5$, and $b$ is the background yields.}
\label{tab:det_config} 
\end{table}

Table~\ref{tab:det_config} shows different detector configurations and the corresponding exclusion limit for $m_{\chi} = 10$ $\textrm{MeV}/\textrm{c}^2$. The default configuration is in the first row, $N_\chi$(relative)=1, $\textrm{N}_\textrm{PE}$(relative)=1, $b=1$, and $n=4$ where ``relative'' means relative to the baseline. 

The improvements we can achieve by extending the duration of data collection or making the detector area larger (adding more bars to each layer) is very modest. Extending the duration or the detector area by a factor of 2 increases $N_\chi$ by the same amount. Making an assumption that the background remains same, the sensitivity is only improved by $10$\% ($2^\textrm{nd}$ row in Table~\ref{tab:det_config}). 	

The impact of $b$ is also small. If $b$ is increased by a factor of $100$, the limit is degraded by $15$\% ($3^\textrm{rd}$ row in Table~\ref{tab:det_config}).  

As the $4^\textrm{th}$ and $5^\textrm{th}$ rows in Table~\ref{tab:det_config} show, the most effective component to enhance sensitivity is $\textrm{N}_\textrm{PE}$. Increasing $\textrm{N}_\textrm{PE}$ by a factor of $2(3)$ improves the sensitivity limit by $25(35)$\%. This can be achieved by using an inorganic crystal with large Z such as BGO or by making the scintillator bars longer. \texttt{GEANT4} simulation shows that a $5\times5\times20$ $\textrm{ cm}^3$ BGO bar provides $N_\gamma=1.2 \times 10^6$, about a factor of $2$ larger than the $5\times5\times80$ $\textrm{ cm}^3$ BC-408 scintillator. Further increasing the length of the bar to $40\textrm{ cm}$, one can achieve $N_\gamma = 1.7 \times 10^6$. However, cost should be considered as well because BGO is less cost-effective than plastic scintillators. 

Though $\textrm{N}_\textrm{PE}$ plays the main role in enhancing sensitivity to sub-millicharged particles, the exclusion limit is still in the range of $\epsilon=1.5-3.0\times10^{-4}$ with the variations considered. This indicates that the sensitivity does not strongly depend on the configuration of the detector. 

With 3 layers the number of detected signal events is larger because $P=(1-e^{-\textrm{N}_\textrm{PE}})^3$ instead of $(1-e^{-\textrm{N}_\textrm{PE}})^4$. However, the background increases as well; the random coincidence rate is approximately $3000$ with 3 layers. Using $n=3$ and $b=3000$, the exclusion limit is slightly improved for $m_\chi<0.2$ $\textrm{GeV}/\textrm{c}^2$ as shown in Fig~\ref{fig:limit}. However, the experience from the milliQan demonstrator~\cite{Ball:2020dnx} gives a lesson that the random coincidence and the correlated backgrounds (induced by either cosmic showers or radiation within the scintillator bars) overlapping with dark current are potentially the major background sources. Adding one more layer (4-layer configuration) helps understanding and rejecting these backgrounds so that the experiment is carried out in a low background environment. This will make the observation of a potential signal more robust. 

\section{Summary}

We studied the feasibility of an experiment searching for millicharged particles using 30 GeV proton fixed-target collisions at J-PARC. With $\textrm{N}_\textrm{POT}=10^{22}$, the experiment provides a sensitivity to $\chi$s with charge down to $\epsilon\simeq3\times10^{-4}$ in $m_\chi<0.2$ $\textrm{GeV}/\textrm{c}^2$ and $\epsilon\simeq1.5\times10^{-3}$ in $m_\chi<1.6$ $\textrm{GeV}/\textrm{c}^2$; this is the regime largely uncovered by the previous experiments. We also explored a few detector designs to achieve an optimal sensitivity to $\chi$s. The photoelectron yields is the main driver, but the sensitivity does not have a strong dependence on the detector configurations in the sub-millicharge regime.

\section*{Acknowledgements}

We thank Tsutomu Mibe,  Yoshiaki Fujii, Takeshi Nakadaira, and Toshifumi Tsukamoto for the useful discussions on the detector site. In particular, we thank Toshifumi Tsukamoto for taking photographs of the Neutrino Monitor building so that we understand the spatial constraints inside the building. We also thank Hong Joo Kim for the discussion on the property of various scintillation materials.

\bibliographystyle{unsrt}
\bibliography{mybibfile}

\end{document}